\documentclass[letterpaper,preprint,prl,aps,superscriptaddress,floatfix]{revtex4}
\usepackage[latin1]{inputenc}
\usepackage{bm}
\usepackage{multirow,amssymb,amsbsy,amsmath}
\usepackage{graphicx}
\usepackage{verbatim}
\usepackage{color}
\makeatletter
\usepackage{pifont}
\makeatother

\usepackage{graphicx}
\usepackage{dcolumn}
\usepackage{bm}
\usepackage{ifthen}
\usepackage{booktabs}
\usepackage{nicefrac}
\usepackage{float}
\usepackage{bbm}
\usepackage{amsmath}

\begin{document}

\title{Experimentally Robust Self-testing for Bipartite and Tripartite Entangled States}
\author{Wen-Hao Zhang}
\author{Geng Chen}
\email{chengeng@ustc.edu.cn}
\author{Xing-Xiang Peng}
\author{Xiang-Jun Ye}
\author{Peng Yin}
\author{Ya Xiao}
\author{Zhi-Bo Hou}
\author{Ze-Di Cheng}
\author{Yu-Chun Wu}
\author{Jin-Shi Xu}
\author{Chuan-Feng Li}
\email{cfli@ustc.edu.cn}
\author{Guang-Can Guo}
\affiliation{CAS Key Laboratory of Quantum Information, University of Science and Technology of China, Hefei, Anhui 230026, China.}
\affiliation{Synergetic Innovation Center of Quantum Information $\&$ Quantum Physics, University of Science and Technology of China, Hefei, Anhui 230026, China.}

\date{\today}
\begin{abstract}
Self-testing refers to a method with which a classical user can certify the state and measurements of quantum systems in a device-independent way.
Especially, the self-testing of entangled states is of great importance in quantum information process. A comprehensible example is that violating the CHSH inequality maximally necessarily implies the bipartite shares a singlet. One essential question in self-testing is that, when one observes a non-maximum violation, how close is the tested state to the target state (which maximally violates certain Bell inequality)? The answer to this question describes the robustness of the used self-testing criterion, which is highly important in a practical sense. Recently, J. Kaniewski predicts two analytic self-testing bounds for bipartite and tripartite systems. In this work, we experimentally investigate these two bounds with high quality two-qubit and three-qubit entanglement sources. The results show that these bounds are valid for various of entangled states we prepared, and thus, we implement robust self-testing processes which improve the previous results significantly.
\end{abstract}

\maketitle 
\bibliographystyle{prsty}
The device-independent science (DI), which is inspired by the requirements for secure quantum information processing, has attracted lots of interests in the past ten years \cite{Scarani2012}. In a DI approach, the only way to study
the system is to perform local measurements and analyze the statistical results. Under the sole assumptions of no-signalling and the validity of quantum theory, it has been shown that it is possible to characterize the quantum devices in quantum key distribution \cite{Acin2006,Acin2007}, randomness
generation \cite{Pironio}, entanglement witness and dimension witness \cite{Gallego} in a DI way. Especially, in some cases, one can
certify uniquely the state and the measurements that are present in the devices, simply by
querying the devices with classical inputs and observing the correlations in the classical outputs. This phenomena is known as the concept of ``self-testing" \cite{Yao2004}. An explicit example is the fact that the maximal violation of the CHSH Bell inequality \cite{Bell1964,Clauser} identifies uniquely the maximally entangled state of two qubits \cite{Popescu1992}, namely ``singlet".

A general conclusion can be made that the violation of a Bell inequality \cite{Brunner2014} reveals the presence of entanglement. Concretely speaking, certain Bell correlations can be reproduced only by performing specific local measurements on a specific entangled state (up to local unitaries), and thus, the observation of such correlations allows one to characterize an unknown source of quantum states, as well as the measurement devices \cite{Gomez2016} or observables \cite{Kaniewski2017,Brunner2018,SUpic2016,McKague2011}, in a DI manner. Recently, criterions to self-test different forms of entangled states, i.e., multi-partite entangled states\cite{Liang}, graph states \cite{McKague}, high-dimensional maximally entangled states \cite{Slofstra,Salavrakos},
non-maximally entangled states of two qubits \cite{Yang,Bamps} or arbitrary pure bipartite states \cite{Scarani2017,Acin2018} are also proposed. The results in these works are limited to ideal scenarios in which the tested states are hundred-percent perfect. However, in practical quantum information processes, the state contained in devices generally deviates from the ideal case in the presence of errors. As a result, we can not observe the maximal quantum violation in self-testing procedures. In such a case, we still want to know how close the tested state is to $\Psi_{A'B'}$ (which maximally violates certain Bell inequality), in other words, how robust the used self-testing criterion is considering realistic errors. Self-testing statements of practically relevant robustness
are of great significance, but few theoretical results are known \cite{Kaniewski2017,Bardyn,Bancal2015}.

Recently, Kaniewski develops a new technique to prove
analytic self-testing statements \cite{Kaniewski2016}. The new method can give rise to a family of operators
to place a lower bound on the spectrum of these operators, and thus, immediately yields a self-testing robustness statement.
The advantage of the new method is that it provides an
explicit construction of the extraction channels in terms of
measurement operators. Previous methods, on the other hand, resort to a
numerical optimization over a wide class of extraction maps and, hence,
do not identify the optimal ones. This distinct advantage makes the given bound flexible to be experimentally implemented.
For singlet self-testing with CHSH inequality, it can give a linear bound which improves on all
the previously known results; for tripartite Greenberger-Horne-Zeilinger (GHZ) state \cite{Greenberger}
self-testing with Mermin inequality \cite{Mermin} it yields the first tight
self-testing statement. In this work, we experimentally investigate these two bounds with high quality two-qubit and three-qubit entanglement sources, which can generate pure and mixed quantum states with adjustable degree of entanglement. The results show that these two bounds are valid for various of entangled states we prepared. Furthermore, by preparing special family of entangled states, we experimentally demonstrate robust self-testing processes which improve the previous results significantly, and thus, our work can be more instructive when apply self-testing to new quantum techniques.

\section{Theoretical Framework}
We describe the self-testing scenario in detail: Alice and Bob share some quantum states in a blackbox-like device, and they want to identify the state through some measurement apparatus (MA). If these MAs can be trusted, they could perform tomography to deduce precisely the form of the shared state. Otherwise, their actions are limited to choosing
the measurement setting and observing the outcome
and hence the only information available to them is the
conditional probability distribution P$(a, b|x, y)$ (i.e., the probability of
observing outputs $a, b$ for inputs $x, y$).

The self-testing statement can be quantified by the extractability of the test state $\rho_{AB}$ to a target state $\Psi_{A'B'}$, which can be defined as
\begin{equation}
\label{extractability}
{\Xi(\rho_{AB}\to\Psi_{A'B'})}:=\max\limits_{\Lambda_A,\Lambda_B} F((\Lambda_A\otimes\Lambda_B(\rho_{AB}),\Psi_{A'B'}),
\end{equation}
where the maximum is taken over all quantum channels (completely positive trace-preserving maps) of the correct input/output dimension and the fidelity is defined as $F(\rho,\sigma)=||\sqrt\rho\sqrt\sigma||_1^2$  ($||...||_1^2$ represents the trace form). In order to test the entanglement characteristics of $\rho_{AB}$, $\Psi_{A'B'}$ is assumed to be a state which achieves the maximal quantum violation. Theoretically, when the maximal quantum violation is observed in a self-testing scenario, the shared unknown state can be mapped to $\Psi_{A'B'}$ with the extractability to be 1. In practical quantum information processes, errors are unavoidable and the robustness can be described by the lowest possible extractability when one observes the violation of (at least) $\beta$ on the Bell inequality $B$, and this quantity can be captured by a function defined as
\begin{equation}
\label{self-testing bounds}
Q_{\Psi,B}(\beta):=\inf\limits_{\rho_{AB}\in S_{B}(\beta)} \Xi(\rho_{AB}\to\Psi_{A'B'}),
\end{equation}
where $S_B(\beta)$ is the set of bipartite states which violate Bell inequality at least with a value $\beta$.

By constructing the local extraction channels $\Lambda_A$ and $\Lambda_B$ from the measurement parameters $a$ and $b$ \cite{Kaniewski2016}, it can be proved
\begin{equation}
\label{CHSH define}
K(a,b)\ge\frac {4+5\sqrt{2}} {16} W_{\alpha,\beta}(a,b)-\frac{1+2\sqrt{2}} {4}.
\end{equation}
The Bell operator $W$ is defined by the parameters $a,b$ and written as $W(a,b)=\sum\limits_{j,k\in\{0,1\}} (-1)^{jk} A_j(a) \otimes B_k(b).$
$K$ is the fidelity operator constructed from $a,b$ and written as $K:=(\Lambda_A(a) \otimes \Lambda_B(b))(\Psi_{A'B'})$ \cite{Kaniewski2016}.

The lower and upper bound can be deduced from Eq. (\ref{CHSH define}) as

\begin{equation}
\label{chsh bound}
\frac {4+5\sqrt{2}} {16}\beta-\frac{1+2\sqrt{2}} {4}\leq Q_{\Psi_{\rm{AB}},B_{\rm{CHSH}}}(\beta_{\rm{CHSH}})< 0.5+0.5*\frac{\beta-2} {2\sqrt{2}-2}
\end{equation}

In a similar way, for three-qubit scenario in which each of the three parties have a binary measurement operator and an extraction channel defined by $a,b,c$. The operator inequality can be written as
\begin{equation}
\label{mermin define}
K(a,b,c)\ge\frac {2+\sqrt{2}} {8} W(a,b,c)-\frac{1}{\sqrt{2}}.
\end{equation}
Surprisingly using Mermin inequality a perfectly tight self-testing bound can be obtained \cite{Kaniewski2016}, as
\begin{equation}
\label{mermin bound}
Q_{\Upsilon_{\rm{ABC}},B_{\textbf{Mermin}}}(\beta)=\frac {1} {2}+\frac{1} {2} \cdot \frac {\beta-2\sqrt{2}} {4-2\sqrt{2}}.
\end{equation}
where $\beta>2\sqrt{2}$ to guarantee a nontrivial fidelity statement.

These two bounds can guarantee higher merit of fidelity when observing a certain value of violation. In other words, previous results only
give a non-trivial bound on the fidelity to the singlet for violations
exceeding 2.37, whereas the new bound brings the threshold down to
2.11.

\section{Experimental Results}
The bounds from
Ref. \cite{Kaniewski2016} exhibit excellent robustness to noise, so they are perfectly
suited to be applied to an actual experiment. This is precisely the
motivation for the current work.

\begin{figure*}[t]
\centering
\includegraphics[width=\textwidth]{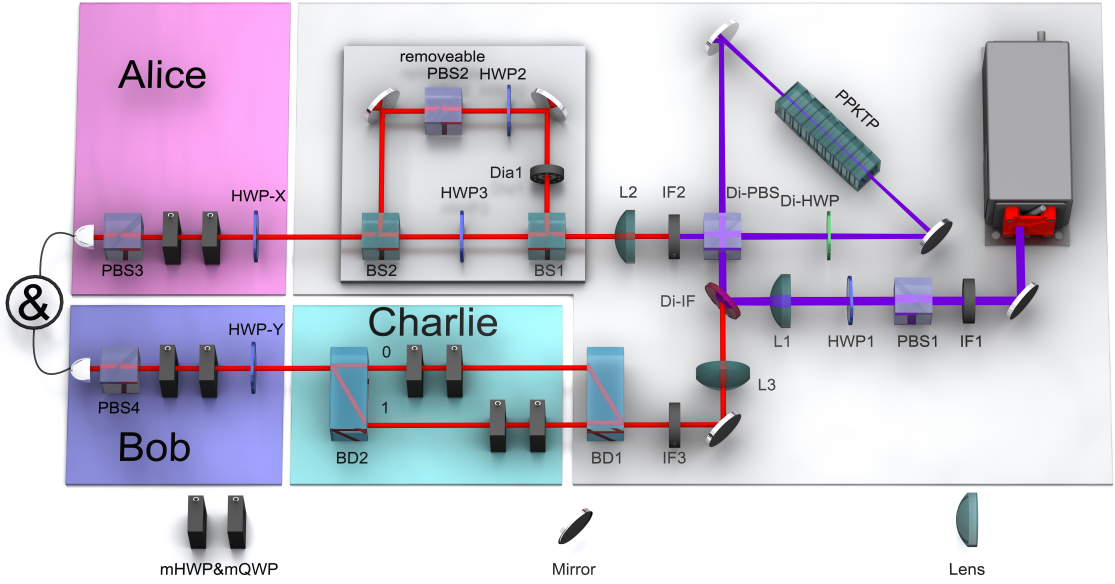}
\caption{\label{Fig1} Experimental setup. A periodically poled KTP (PPKTP) nonlinear crystal placed inside a phase-stable Sagnac interferometer (SI) is pumped by a single mode 405.4nm laser to produce polarization-entangled photon pairs at 810.8nm. Bandpass filters at 810nm and long-pass filters are used to block the pump light. One photon will pass through an unbalanced(enough to make two paths light fully decoherent) Mach-Zehnder interferometer (MZI) constructed by Beam Splitters BS1 and BS2 when mixed states need to be generated. The other photon will pass through a balanced MZI constructed by Beam displacers BD1 and BD2 when we need to generate tripartite states. One polarization beam splitter (PBS),  motorized half-wave plate (mHWP) and motorized quarter-wave plate (mQWP) are used to perform projection measurements on each qubit.
 }
\end{figure*}

The above two analytic self-testing bounds in inequalities (\ref{chsh bound}) and (\ref{mermin bound}) can be experimentally studied by the setup in Fig. 1.
The setup consists of two main parts. One part is responsible for the generation of polarization entangled photon pairs, and the other part is in charge of generating various kinds of, pure or mixed, two-qubit and three-qubit states. In the first part, a 405.4 nm continuous-wave diode laser with polarization set by a half-wave plate (HWP) is used to pump a 4 mm-long PPKTP crystal inside an SI to generate polarization-entangled photons \cite{Fedrizzi}. The photon pairs are in the state $\cos\theta|HH\rangle+\sin\theta|VV\rangle$ ($H$ and $V$ denote the horizontal and vertical polarized components, respectively) and $\theta$ is controlled by the pumping polarization. The visibility of the maximally entangled state is larger than 0.985. These photon pairs are then sent into the second part by two single mode fibers, and the polarization is maintained by two HWPs before and after each fiber. In the second part, on Alice's side an extra delay circuit is introduced by two BSes, which can be used to generate mixed states. On Bob's side, a MZI induces an additional path mode on the passing photon, which is assigned to be a third qubit, namely, Charlie.

For the generation of two-qubit states, Charlie does nothing and can be neglected. Alice and Bob can perform measurement operators $K$ and $W$ with various $(a,b)$, and we want to verify that inequality (\ref{chsh bound}) cannot be violated for all the tested scenarios. $W(a,b)$ can be decomposed into ${A0,A1,B0,B1}$ to be realized by a QWP-HWP array, which are in the form of
\begin{equation}
\label{basis}
\begin{split}
A(B)_r &= cos\,a(b)\;\sigma_x+(-1)^r sin\,a(b) \;\sigma_z
\end{split}
\end{equation}
with $r\in{0,1}$. As for $K(a,b)$, it is made up by mixtures of Pauli matrices $\sigma_{i}\otimes\sigma_{i}$ ($i\in{1,2,3,4}$), of which the expected value can be directly measured, thereby $\langle K(a,b)\rangle$.

\begin{figure}[tb]
\centering
\includegraphics[width=1\textwidth]{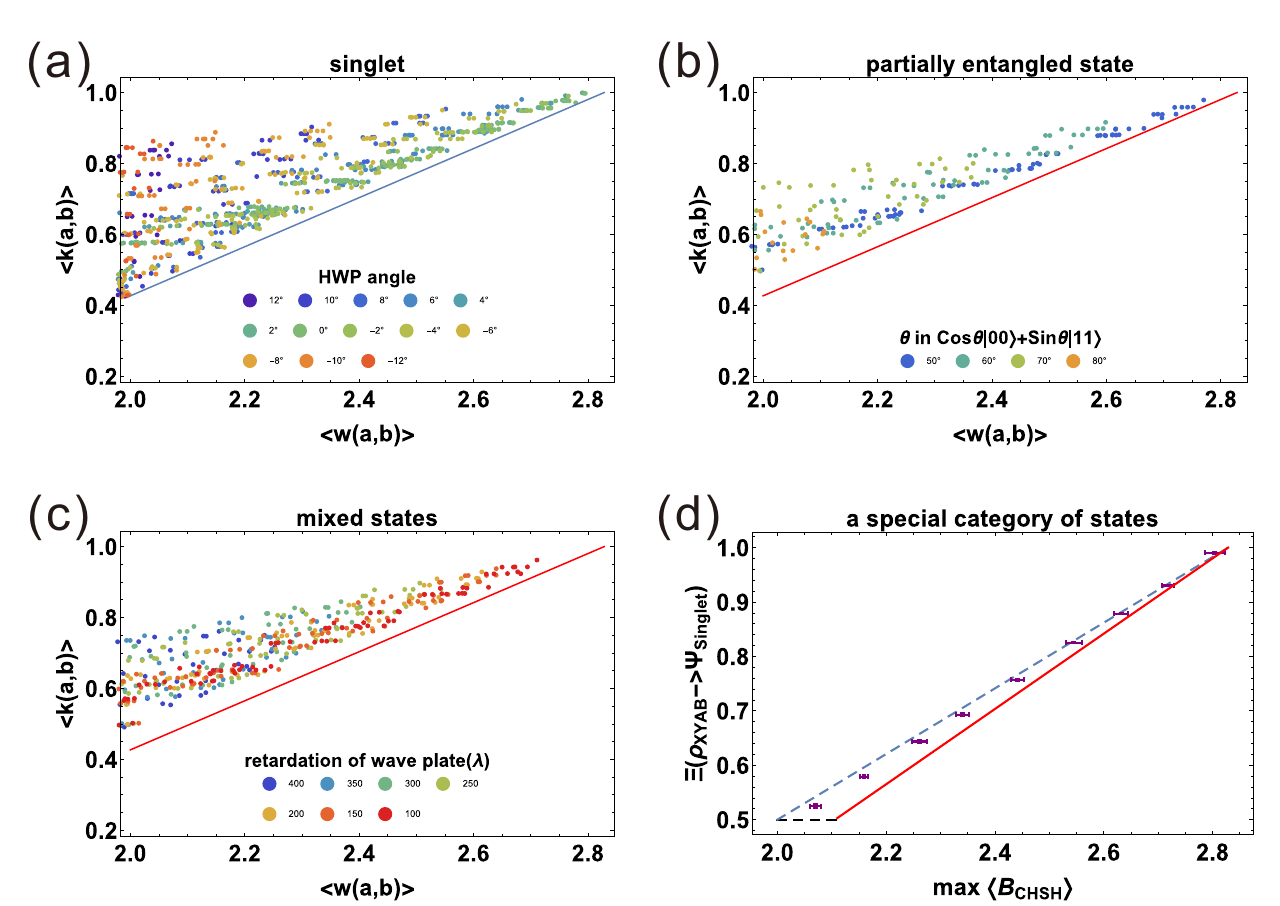}
\caption{\label{Fig2}  Experimental test of operator inequality (\ref{CHSH define}) with (a) different singlet states obtained by rotating HWP3 to the angles shown in the figure; (b)partially entangled states characterized by the value of $\theta$ shown in the figure; (c)mixed states by inserting waveplates of different levels of retardation. With random sampled $\{a,b\}$, all the data points are above the given bound. (d) Experimental self-testing of a special category of example states $\rho_{XYAB}$, which are composed by four components registered by $XY (X,Y=0,1)$. Changing the weight of the maximally entangled component $\tau_{AB}^{11}$, a series of states are self-tested and all the data points fall in the narrow area between the given lower and upper bounds. Because the upper bound is very close to the lower bound, it can be concluded from these results that the lower bound is nearly optimal. }
\end{figure}

The results for different singlets, partially entangled and mixed bipartite states are shown in Fig. 2(a)-(c). It is clear for all the tested states and different values of $(a,b)$, the data points are all above the red line, which exactly satisfies the inequality (\ref{CHSH define}). As a result, the left side of inequality (\ref{chsh bound}) is valid for all the tested scenarios. In order to verify that the given bound is nearly optimal, which allows a robust self-testing process, a special category of states are investigated by measuring the relation between the maximum Bell violation $\beta$ and the extractability $\Xi$. We experimentally prepared a family of states as
\begin{equation}
\label{example state}
 \rho_{\rm{XYAB}}=\sum_{xy} p_{xy}|x\rangle\langle x|_X \otimes |y\rangle\langle y|_Y \otimes \tau_{AB}^{XY},
\end{equation}
where $p_{11}= \nu$ and $p_{00}=p_{01}=p_{10}= (1-\nu)/3$ for $\nu\in(0,1)$. Specially $\tau_{AB}^{XY}$ shared by Alice(A) and Bob(B) with respective classical register qubit $X, Y$ are $\tau_{AB}^{00} = \tau_{AB}^{01}=\tau_{AB}^{10}=\frac{1} {4}(\mathbbm{1}\otimes\mathbbm{1}+\sigma_x\otimes\sigma_x)$ and $\tau_{AB}^{11} = \frac{1} {4}(\mathbbm{1}\otimes\mathbbm{1}+\frac{1} {\sqrt{2}}[\sigma_x\otimes\sigma_x+\sigma_x\otimes\sigma_z+\sigma_z\otimes\sigma_x-\sigma_z\otimes\sigma_z]+\sigma_y\otimes\sigma_y)$.

 In experiment, $\tau_{AB}^{11}$ can be prepared by closing the diaphragm (Dia1) and rotating HWP3 to be $33.75^{\circ}$. On the other hand, the other three components, $\tau_{AB}^{00}$, $\tau_{AB}^{10}$ and $\tau_{AB}^{01}$ can be generated by opening the Dia1, meanwhile removing HWP2 and PBS2 and setting HWP3 to be $45^{\circ}$. The four states can simultaneously achieve the optimal value of the CHSH expression, which equals $2\sqrt{2}$ for $\tau_{AB}^{11}$ and 2 for the
remaining three states. Here we chose $\tau_{AB}^{11}$, $\tau_{AB}^{10}$,  $\tau_{AB}^{01}$ to have their maximal fidelity while $\tau_{AB}^{00}$ can not obtain the maximal fidelity (usually it's not the optimal strategy, however we use it as an example to reveal the fact that these four states can not achieve maximal fidelity with $\Psi_{A'B'}$.)

Obviously $\tau_{AB}^{11}$ is a pure maximally entangled state and chosen to be the state which Alice and Bob try to extract. For $X=Y=1$ Alice and Bob do nothing and the fidelity to $\tau_{AB}^{11}$ can be 1, while the extraction channels (realized by HWP-X and HWP-Y) can only achieve the maximum fidelity 1/2 for two of the other three components. As a result, $\rho_{\rm{XYAB}}$ can not reach the upper bound for any value of $\nu\in(0,1)$, as shown in Fig. 2(d). All the points are below the upper bound (dotted line), which is very close to the lower bound (red line). Hence, we experimentally prove that the given robustness bound to self-test bipartite states is valid for all the states we tested. Furthermore, we show that the given bound is nearly optimal to implement a robust self-testing procedure, by identifying the states falling in the narrow band between the upper and lower bounds.

By introducing Charlie we can generate various of tripartite entangled states and measure the corresponding $\langle K(a,b,c) \rangle$ and $\langle W(a,b,c) \rangle$. In experiment, the simulated tripartite states are generated by introducing the spatial mode of one photon, which is realized by BD1 and BD2 in Fig. (1). All these data points are well above the lower bound described by inequality (\ref{mermin define}), as shown in Fig. 3(a).

\begin{figure}[tb]
\centering
\includegraphics[width=1\textwidth]{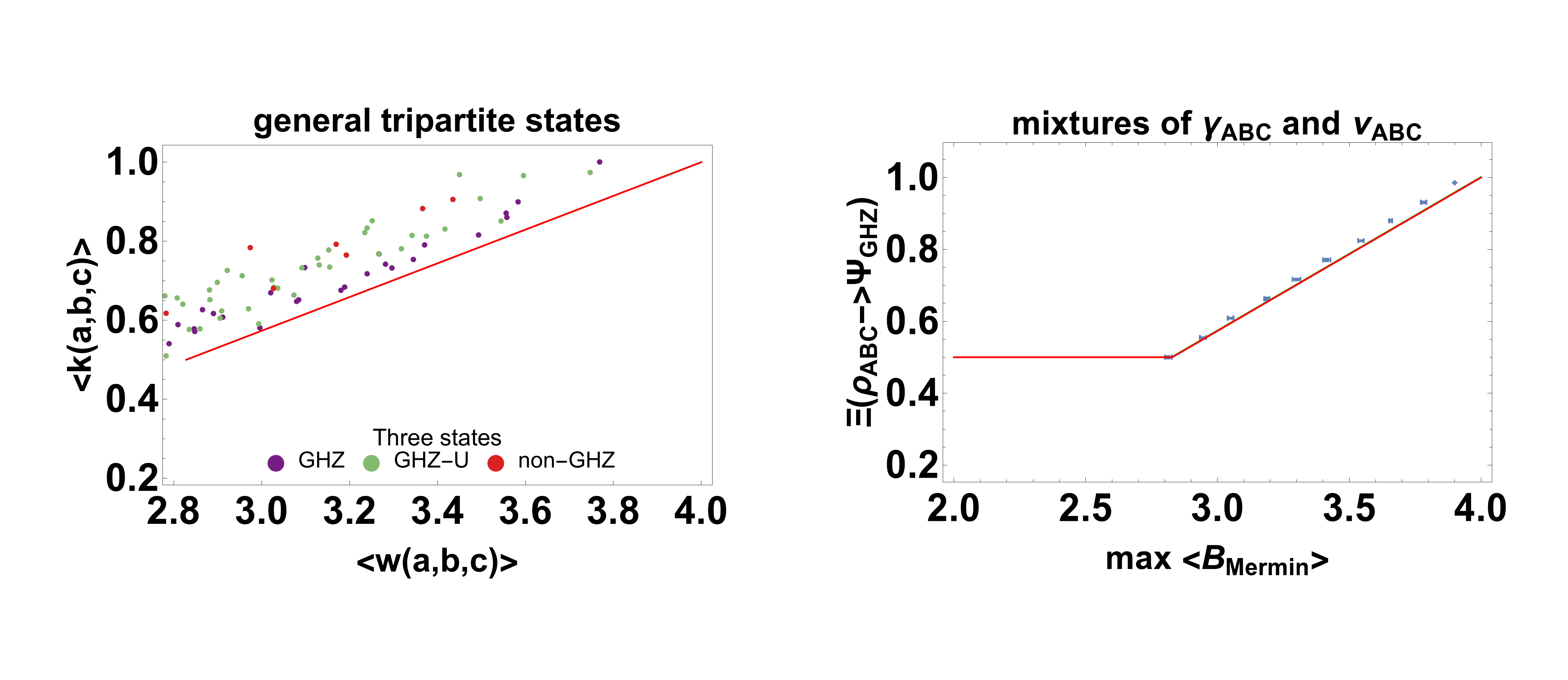}
\caption{\label{Fig3} Experimental test of the self-testing bound for tripartite scenarios. (a) For several typical tripartite entangled states, both the mean values of fidelity operator $K$ and Mermin operator $W$ are measured, with random sampling on $(a,b,c)$. All the data points are above the bound (red line) given by operator inequality (\ref{mermin define}). (b) For mixtures of $\Upsilon_{ABC}$ and $\nu_{ABC}$ with different weights, the maximum violations of Mermin inequality and corresponding extractabilities are measured. All the dots almost fall on the lower bound (red line), indicating the given bound is tight.}
\end{figure}

 For tripartite self-testing, we prepare a family of states which are mixtures of a GHZ state $\Upsilon_{\rm{ABC}}$ and $\nu_{\rm{ABC}}$, where $\nu_{\rm{ABC}}=|\Phi\rangle_{\rm{AB}}|0\rangle_{\rm{C}}$ with $\Phi_{\rm{AB}}$ to be equivalent to the singlet state (up to local unitary). By changing the mixture weight, a series of mixed states are generated with a fidelity $\simeq 0.98$ and the measured results are shown in Fig. 3(b). The data points are almost fall on the lower bound (red line) considering the error bars, indicating the given bound for tripartite scenario is provably tight to perform a robust self-testing procedure.

Device-independent certifications require no-signalling constraints on the devices \cite{Hensen2015}, which can be tested through the influence to one side from the measurements of all the other sides \cite{Brunner2014}. For qubit systems where an operator has a binary outcome, no-signalling requires the mean value of an observable on one side is unaltered with different measurement settings of all the other sides. For example, in case of a tripartite state, the measurement settings are $x$ and $y, z$ on Alice, Bob and Charlie sides respectively, with corresponding outcomes to be $a,b,c$. As a result, the mean value of $\langle A_{x}\rangle$ can be calculated as by summing all the possibilities on $b$ and $c$. No-signalling constraint requires that $\langle A_{x}\rangle$ is a constant when ($y,z$) is changed to be ($y',z'$), which can be written as
\begin{align}
& \langle A_{x}\rangle=\sum_{b=0}^{1}\sum_{c=0}^{1}(P(a=1,b,c|x,y,z)-P(a=-1,b,c|x,y,z)) \\
& =\sum_{b=0}^{1}\sum_{c=0}^{1}(P(a=1,b,c|x,y',z')-P(a=-1,b,c|x,y',z'))\quad for\quad all\quad x,y,y',z,z'
\end{align}
Similar requirements need to be satisfied for all the joint measurement settings appearing in CHSH and Mermin Bell inequalities, and the results are shown in Fig. (4). When the tested states are maximally entangled singlet and GHZ state, the mean values of an arbitrary observable are approximately identical for different joint measurement settings appearing in the inequalities, which exactly satisfies the no-signalling constraint.
\begin{figure}[tb]
\centering
\includegraphics[width=1\textwidth]{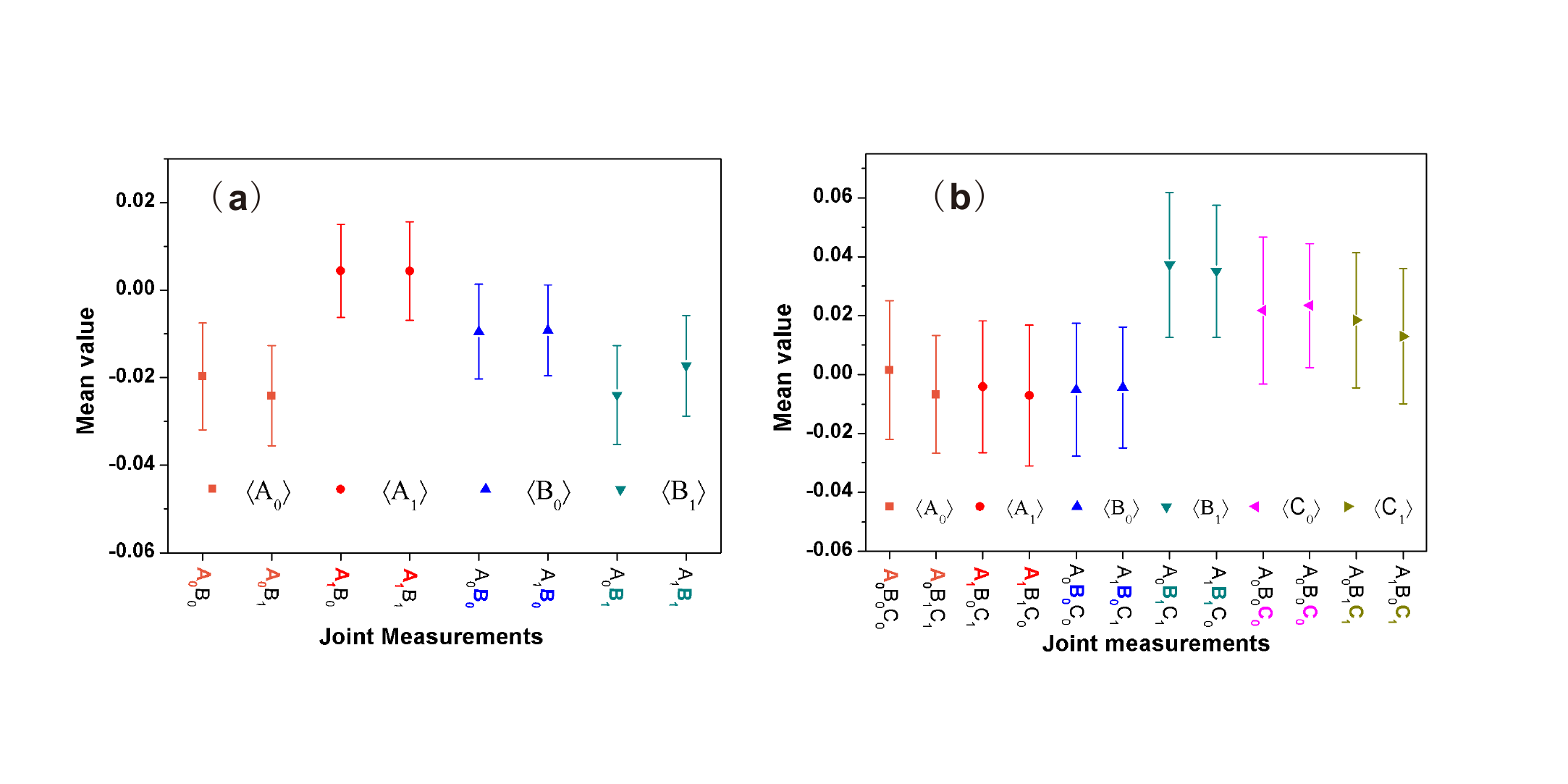}
\caption{\label{Fig4} No-signalling tests for measuring the (a) CHSH inequality with singlet state and (b) Mermin inequality with GHZ state. The mean value of each operator is invariant within one standard deviation, which confirms the no-signalling characteristics of the used devices.
 }
\end{figure}

Although the scientific community pursues complete self-testing criterions with which one can characterize uniquely (up to local isometries) the state and measurements contained in quantum devices, self-testing robustness statement is of significant importance from a practical point of view. Similar statements are mainly inspired by the fact that the realistic states may slightly deviate from the ideally entangled states and become partially mixed . In these cases, the robustness statements are able to provide a quantitative description of the entanglement characteristics about the tested states. As a specific example, one may not much care for the concrete form of the shared states contained in the entanglement sources, while only seeks for a guarantee about the quality of entanglement. In this scenario, simply by querying the devices with classical inputs and observing the correlations in the classical outputs, one can immediately obtain a minimum fidelity to the ideal state according to the robustness bound. The two (nearly) optimal analytic robustness bounds, which are applicable for bipartite and tripartite systems are tested in this experiment. The results clearly confirm the validity of these two bounds, thereby robust self-testing processes are achieved in this experiment. Our work is instructive for practical self-testing tasks, such as placing a lower bound on the distillable entanglement of the unknown state.

{\bf  Acknowledgments}

 This work was supported by the National Key Research and Development Program of China (Nos. 2017YFA0304100, 2016YFA0302700), National Natural Science Foundation of China (Grant Nos. 61327901, 11774335, 91536219), Key Research Program of Frontier Sciences, CAS (No. QYZDY-SSW-SLH003), the Fundamental Research Funds for the Central Universities (Grant No. WK2030020019, WK2470000026).

\bibliographystyle{}

\end{document}